\documentclass[a4paper]{jpconf}
\usepackage{graphicx}
\usepackage{braket}
\usepackage{xcolor}
\usepackage{amsmath}
\begin{document}
\title{Quantum simulation of oscillating neutrinos}
\author{Abhishek Kumar Jha, Akshay Chatla and Bindu A. Bambah}

\address{School of Physics, University of Hyderabad, Hyderabad-500046, India}

\ead{abhiecc.jha@gmail.com, chatlaakshay@gmail.com, bbambah@gmail.com}

\begin{abstract}
{\small Two and three flavor oscillating neutrinos are  shown  to exhibit the properties bipartite and tripartite quantum entanglement. The two and three flavor neutrinos are mapped to qubit states used in quantum information theory. Such quantum bits of the neutrino state  can be encoded on a IBMQ computer using quantum computing as a tool. We show the implementation of entanglement in the two neutrino system on the IBM quantum processor.}
\end{abstract}
\section{Introduction}
In the three flavor neutrino oscillation, the neutrino flavor states $\ket{\nu_{\alpha}}$ ($\alpha=e,\mu,\tau$) are linear superposition of mass eigenstates $\ket{\nu_j}$ ($j=1,2,3 $): $\ket{\nu_{\alpha}}=\Sigma_j U_{\alpha j}\ket{\nu_j}$,  where $U_{\alpha j}$ are the elements of the lepton mixing matrix known as PMNS (Pontecorvo-Maki-Nakagawa-Sakita) matrix such that 
\begin{small}
 \begin{equation}
 \begin{pmatrix}
 \ket{\nu_e}\\
 \ket{\nu_\mu}\\
 \ket{\nu_\tau}\\
 \end{pmatrix}=\begin{pmatrix}
 U_{e1} & U_{e2} & U_{e3}\\
 U_{\mu 1} & U_{\mu 2} & U_{\mu 3}\\
 U_{\tau 1} & U_{\tau 2} & U_{\tau 3}\\
 \end{pmatrix}
 \begin{pmatrix}
 \ket{\nu_1}\\
 \ket{\nu_2}\\
 \ket{\nu_3}\\
 \end{pmatrix}.
 \end{equation}
\end{small}
 The time evolution follows $\ket{\nu_{\alpha}(t)}=\Sigma_j e^{-iE_j t} U_{\alpha j}\ket{\nu_j}$, where $E_j$ is the energy associated with the mass eigenstates $\ket{\nu_j}$. This is a superposition state. Therefore, we expect quantum entanglement \cite{Jha:2020hyh,Blasone:2007vw}. Entanglement of quantum states imply that studies of neutrino states can be addressed using quantum information science techniques. Recently, the work of Arg${\ddot{u}}$elles and Jones \cite{Arguelles:2019phs} has  rekindled interest in the quantum simulation of the neutrino oscillation on IBMQ processors. Earlier, the simulation of neutrino oscillations was analysed using quantum walks \cite{DiMolfetta:2016gzc}. Our paper outlines the various quantum information aspects of neutrinos i.e., the entanglement of neutrinos and how to implement it on a IBM quantum processor. Each flavor state ($\alpha=e,\mu,\tau$) at t=0 is mapped to qbit like states: $
  \ket{\nu_{e}}=\ket{1}_e \otimes \ket{0}_\mu \otimes \ket{0}_\tau \equiv \ket{100}_e$ , $\ket{\nu_{\mu}}=\ket{0}_e \otimes \ket{1}_\mu  \otimes \ket{0}_\tau \equiv \ket{010}_\mu$ and $
 \ket{\nu_{\tau}}=\ket{0}_e \otimes \ket{0}_\mu  \otimes \ket{1}_\tau \equiv \ket{001}_\tau.$

 \section{Bi-partite entanglement in two-flavor neutrino oscillations}

 First we consider two flavor oscillations. The results can easily be generalised to three flavors, which we will do so in the conclusion. In two-flavor ($\nu_\alpha \rightarrow \nu_\beta$) mixing, $\ket{\nu_e(t)}=\tilde{U}_{ee}(t)\ket{10}_e+\tilde{U}_{e\mu}(t)\ket{01}_\mu$, where $\ket{10}$ and $\ket{01}$ are two-qubit states \cite{Blasone:2007vw}:
 $ \ket{\nu_{e}}=\ket{1}_e \otimes \ket{0}_\mu \equiv \ket{10}_e,$ and   $\ket{\nu_{\mu}}=\ket{0}_e \otimes \ket{1}_\mu  \equiv \ket{01}_\mu.$ 
The probability that a neutrino originally of flavor $\alpha$ will later be observed as having flavor $\beta$ is 
$P_{\alpha\rightarrow\beta} = \left|\left\langle\left. \nu_\beta(L) \right| \nu_\alpha \right\rangle \right|^2 = \left|\sum_i U_{\alpha i}^* U_{\beta i}e^{-i\frac{m_i^2 L}{2E}}\right|^2.$
The appearance ($P_a$) and disappearance ($P_d$) probabilities are typically functions of $L/E$ as neutrino masses are very small so in the ultrarelativistic limit in natural units ($c=1) $  $L\approx t$. 
For the two neutrino case, for electron and muon neutrino oscilations, the mixing matrix is  $R(\theta) = \begin{pmatrix} \cos\theta & -\sin\theta \\ \sin\theta & \cos\theta \end{pmatrix}.$
 The probability of changing flavor in a two neutrino system is 
$P_{\alpha\rightarrow\beta , \alpha\neq\beta}=|\tilde{U}_{\alpha \beta}(t)|^2 = 4\,{\sin^2 \theta}\, {\cos^2 \theta} \sin^2\left(\dfrac{\Delta m^2 t}{4E}\right) $
called appearance probabilty $P_a$ and  the probabilty of not changing flavor is 
$P_{d}=|\tilde{U}_{\alpha \alpha}(t)|^2={\cos^4 \theta}+ {\sin^4 \theta }+2 \sin^2\theta\cos^2\theta \cos \left(\frac{\Delta m^2 t}{2E}\right).$  Here we take $\alpha=e$ and $\beta=\mu$. The density matrix for the two neutrino system is
 \begin{equation}
 \rho^{e\mu} (t)=\begin{pmatrix}
 0 & 0 & 0 & 0\\
 0 & \vert{\tilde{U}_{ee}(t)}\vert^2 & \tilde{U}_{ee}(t) \tilde{U}_{e\mu}^*(t) & 0\\
 0 &  \tilde{U}_{e\mu}(t)\tilde{U}_{ee}^*(t) & \vert{\tilde{U}_{e\mu}(t)}\vert^2 & 0 \\
 0 & 0 & 0 & 0\\
  \end{pmatrix}.
 \end{equation} 

\begin{figure}
\begin{center}
     \includegraphics[width=1.0\linewidth]{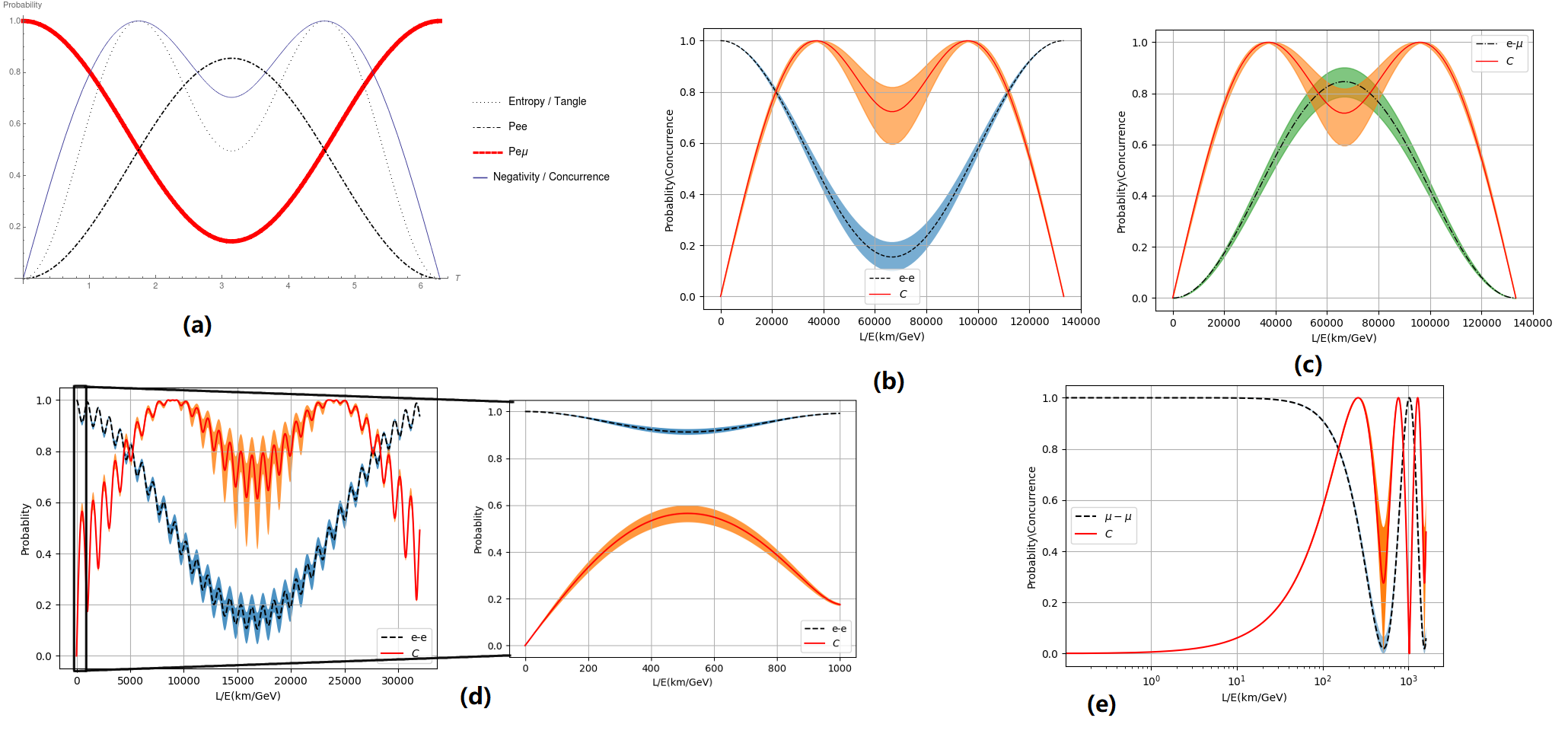} 
  \caption{{\textbf{(a)}  The time evolution of the various measures of entanglement compared to the oscillation probabilities in a typical reactor experiment. $\tau_{e\mu}$ and $S_{e\mu}$ ( dotted line), $N_{e\mu}$ and $C_{e\mu}$ (solid line), $P_{ee}$ (dotted dash line), and $P_{e\mu}$ (dashed line) as functions of the scaled time $T\equiv\Delta{m}^2t/2E$ with the experimental value $\sin^2\theta=0.310$, where $\theta$= mixing angle \cite{Esteban:2018azc}. \textbf{(b)} The blue band shows the $\nu_e$  disappearance probability (Black, dashed line) and the orange band shows the concurrence (Red, solid line) \cite{Esteban:2020cvm}.  \textbf{(c)} The green band shows the $\nu_e$ appearance probability (Black, dash dotted line) and the orange band shows the concurrence (Red, solid line) \cite{Esteban:2020cvm}.  \textbf{(d)} The blue band shows the short range $\nu_e$ disappearance probability (Black, dashed line) and the orange band shows concurrence (Red, solid line), using the Daya Bay experimental data \cite{An:2015rpe}. \textbf{(e)} The blue band represents the long-range survival probability $\nu_\mu\rightarrow\nu_\mu$ (Black, dashed line) and it compared with orange band which gives concurrence (Red, solid line), using the Minos experimental data \cite{Sousa:2015bxa}.}}
\end{center}
\end{figure}

From the density matrix we can extract two measures of entanglement. First, we take the partial transpose of the density matrix and determine the eigenvalues. If the eigenvalues are  all positive, then the state is unentangled. A negative eigenvalue implies entanglement. Taking partial transpose over the muon mode in the neutrino  density matrix we find that that one of the eigenvalues of $\rho^{T_{\mu}}(t)$  is negative and equal to $\lambda_4=-\sqrt{P_aP_d}$, therefore, we can say that under PPT (Positive Partial Transpose) criterion the time evolved electron neutrino flavor state is an entangled state \cite{DARIUSZ:2012}. Another measure of entanglement called  negativity (N) is quantified by the trace norm of the partial transpose of the density matrix $N=Tr\sqrt{\rho^{T_\mu}(t)\rho^{T^{\dagger}_{\mu}}(t)}-1=2|\sum_i\lambda_i|$, where $\lambda_i\leq 0$ are the negative eigenvalues of partial transposition $\rho^{T_\mu}(t)$, for maximally entangled state, $N_{e\mu}$ has to be positive. In this case $ N_{e\mu}=2\sqrt{P_a P_d}> 0$, so this  necessary and sufficient entanglement  condition is  also satisfied in the two neutrino system \cite{OU:2007}. We also find the linear entropy : $S({\rho^{e\mu}(t)})=1-Tr({\rho^{e\mu}(t)})^2$ which is a lower approximation of von-Neumann entropy. There are more robust measures of entanglement such as the  concurrence (C) and tangle ($\tau$), which  are calculated  using the spin-flipped density matrix, $\tilde{\rho}^{e\mu}(t)=(\sigma_y\otimes\sigma_y){\rho^*}^{e\mu}(t)(\sigma_y\otimes\sigma_y)$ where $\sigma_x$ and  $\sigma_y$ are Pauli matrices.  The  concurrence  $C(\rho^{e\mu}(t))\equiv [max(\mu_1 -\mu_2 -\mu_3-\mu_4 ,0)]$, where $\mu_1,...,\mu_4$ are the eigenvalues of the spin-flipped matrix  and  the tangle $\tau(\rho^{e\mu}(t))\equiv [max(\mu_1 -\mu_2 -\mu_3-\mu_4 ,0)]^2$ \cite{OU:2007, Coffman:1999jd}.
 All these entanglement measures are related as \cite{Jha:2020hyh} $\tau_{e\mu}=C^{2}_{e\mu}=N^2_{e\mu}=S_{e\mu}=4P_a P_d.$ Therefore the two flavor neutrino oscillations is a bi-partite entangled system of two qubit pure states. In Fig1(a). we plot the entanglement measures vs scaled time $T\equiv\Delta{m}^2t/2E$ for the time evolved electron neutrino flavor state, where disappearance $P_{d}$ and appearance $P_{a}$ probablities are also shown for comparison \cite{Esteban:2018azc}. 
In order to understand the entanglement behavior of two flavor neutrino oscillations more precisely, the concurrence plot for an initial electron flavor neutrino $\nu_e$ in terms of disappearance $P_{ee}$ and appearance $P_{e\mu}$ probablities with ratio $L/E(Km/GeV)$ changing are shown in Fig1(b) and Fig1(c), respectively \cite{Esteban:2020cvm}. We find that when the disappearance probablity $P_{ee}$ is minimum and the appearance $P_{e\mu}$ probablity is maximum, the concurrence is minimum which means that the particle is disentangled. In Fig.1(d) and Fig.1(e), we also shown the concurrence vary with ratio $L/E$ changing for the short range $\nu_e\rightarrow\nu_e$ and long-range $\nu_\mu\rightarrow\nu_\mu$ disappearance probablities using the Daya Bay and Minos experimental data, respectively \cite{An:2015rpe, Sousa:2015bxa}.

\section{Quantum computer circuit to simulate bi-partite entanglement and concurrence  in two flavor neutrino oscillations on an IBMQ platform}
The universal quantum gate $U3$  in quantum information theory is \cite{Arguelles:2019phs}
\begin{equation}
U3(\Phi,\psi,\lambda)=\begin{pmatrix}
cos\frac{\Phi}{2} & -sin\frac{\Phi}{2}e^{i\lambda}\\
sin\frac{\Phi}{2}e^{i\psi} & cos\frac{\Phi}{2}e^{i(\lambda+\psi)}\\
\end{pmatrix}.
\end{equation}
 $ U3$ can be identified with the SU(2) rotation matrix  $R(\theta)=\begin{pmatrix}
cos\theta & -sin\theta\\
sin\theta & cos\theta\\
\end{pmatrix}$  which is the mixing matrix for two neutrinos.
This is done  by rephasing the  neutrino states 
\begin{eqnarray}
&{\ket{\nu_\mu}\rightarrow e^{-i\psi}\ket{\nu_\mu}}&\nonumber\\
&{\ket{\nu_2}\rightarrow e^{i\lambda}\ket{\nu_2}}.&
\end{eqnarray} 
Without loss of generality, we can set the parameter value $\psi=0$ and $\lambda=0$ in U3 gate.
In quantum optics, the action of quantum mechanical beam splitter is given by $SU(2)$ rotation matrix $R(\theta)$ which performs exactly the same transformation on photons as the neutrino mixing matrix does. Thus, the action of $U3(2\theta,0,0)$ is akin to a beam splitter transformation and it can transfer each bit into the  superposition states  : 
\begin{eqnarray}
& {U3(2\theta,0,0)\ket{0}}\rightarrow{\tilde{U}_{e e}\ket{0}+\tilde{U}_{e\mu}\ket{1}}, \nonumber& \\
& {U3(2\theta,0,0)\ket{1}}\rightarrow{\tilde{U}_{e e}\ket{1}+\tilde{U}_{e\mu}\ket{0}}. &
\end{eqnarray}
The time-evolution operation can be identified as an S-gate given by
\begin{equation}
S(\psi)=\begin{pmatrix}
1 & 0\\
0 & e^{i\psi}\\
\end{pmatrix}=U1(t).
\end{equation}
For the two $\nu$ system the entire time evolution circuit is given by combinations of $U3$ and $S$ gates \cite{Arguelles:2019phs},
\begin{small}
 \begin{eqnarray}
&{\begin{pmatrix}
\ket{\nu_e(t)}\\
\ket{\nu_\mu(t)}\\
\end{pmatrix}=\begin{pmatrix}
cos\theta & sin\theta\\
-sin\theta & cos\theta\\
\end{pmatrix} 
\begin{pmatrix}
1 & 0\\
0 & e^{i\psi}\\
\end{pmatrix}
\begin{pmatrix}
cos\theta & -sin\theta\\
sin\theta & cos\theta\\
\end{pmatrix}
\begin{pmatrix}
\nu_e(0)\\
\nu_\mu(0)
\end{pmatrix} \equiv\begin{pmatrix}
\tilde{U}_{ee}(t) & \tilde{U}_{e\mu}(t)\\
\tilde{U}_{\mu e}(t) & \tilde{U}_{\mu\mu}(t)\\
\end{pmatrix}\begin{pmatrix}
\nu_e(0)\\
\nu_\mu(0)\\
\end{pmatrix}},  & 
\end{eqnarray}
\end{small}
where $\psi=\frac{\Delta m^2 t}{2E}$. The quantum circuit embodying two-flavor neutrino oscillations contains following steps: Prepare flavor state $\rightarrow$ Rotate to mass basis  $\rightarrow$ Time evolution $\rightarrow$ Rotate from mass basis $\rightarrow$ Measure the state.

\begin{figure}
\begin{center}
      \includegraphics[width=1.0\linewidth]{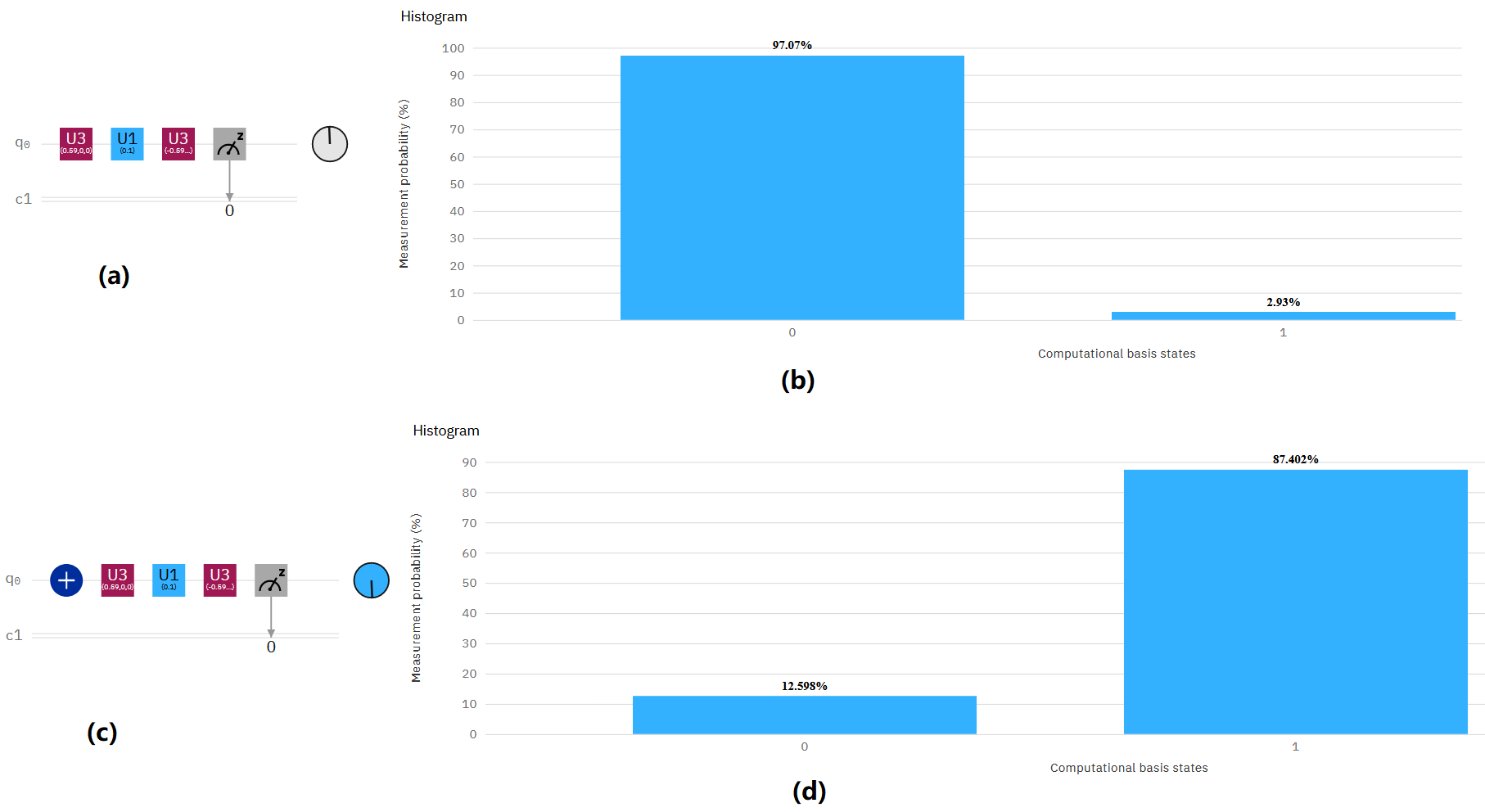} 
     \caption{ Two flavor neutrino oscillations encoded in one qubit mode system on IBMQ platform. \textbf{(a)} and \textbf{(b)} shows circuit diagram and histogram plot (disappearance probablities in percentage) for $P_{e\rightarrow e}$: ($\ket{\nu_e(t)}=\tilde{U}_{ee}(t)\ket{0}_{e}+\tilde{U}_{e\mu}(t)\ket{1}_{\mu}$). \textbf{(c)} and \textbf{(d)} shows circuit diagram and histogram plot (appearance probablities in percentage) for  $P_{\mu\rightarrow e}$: $\ket{\nu_\mu(t)}=\tilde{U}_{\mu e}(t)\ket{0}_{e}+\tilde{U}_{\mu\mu}(t)\ket{1}_{\mu}.$}
 \end{center}    
 \end{figure}     
    
The gate arrangement of two flavor neutrino oscillations  in one qubit mode system can be expressed as (see Fig.2):
  \begin{eqnarray}
&{\ket{\nu_e(t)}}={U3(2\theta,0,0)U1(t)U3(-2\theta,0,0)\ket{0}}&\nonumber\\
&{\ket{\nu_\mu(t)}}={U3(2\theta,0,0)U1(t)U3(-2\theta,0,0)X\ket{0}}.&
\end{eqnarray}
Now, we go onto the two qbit system. This is required because the doubling of states is required to calculate measures such as concurrence and tangle , which are calculated from the density matrix.  We first prepare a quantum circuit of bi-partite electron neutrino state in the linear superposition of mass mode basis (Fig3(a)) 
  \begin{eqnarray}
 &{\ket{\nu_e(0)}=CNOT_{12}[U3(-2\theta,0,0)\ket{0}_1\otimes X\ket{0}_2] \rightarrow{CNOT_{12}[(\tilde{U}_{ee}\ket{1}_1+\tilde{U}_{e\mu}\ket{0}_1)\otimes\ket{1}_2}]}&\nonumber \\
 &{\rightarrow \tilde{U}_{ee}\ket{10}_1+\tilde{U}_{e\mu}\ket{01}_2}.&    
\end{eqnarray}   
The $CNOT_{12}$ gate is defined such that  if the control qubit (first (1) qubit) is in the state $\ket{0}$ the target qubit (second (2) qubit ) is not affected, conversely if the control qubit in the state $\ket{1}$, the target is flipped. 
The gate arrangement of the time evolved electron flavor neutrino state in the two qubit flavor mode system is (Fig3(b))
\begin{equation}
\ket{\nu_e(t)}
=CNOT_{12}[U3(2\theta,0,0)U1(t)U3(-2\theta,0,0)\ket{0}_1\otimes X\ket{0}_2] \rightarrow \tilde{U}_{ee}(t)\ket{10}_e+\tilde{U}_{e\mu}(t)\ket{01}_\mu.
\end{equation}
The gate arrangement of the time evolved muon flavor neutrino state in the two qubit flavor mode system is (Fig3(d))
\begin{equation}
 \ket{\nu_\mu(t)}
={CNOT_{12}[U3(2\theta,0,0)U1(t)}U3(-2\theta,0,0)X\ket{0}_1\otimes X\ket{0}_2] \rightarrow \tilde{U}_{\mu e}(t)\ket{10}_e+\tilde{U}_{\mu\mu}(t)\ket{01}_\mu .
    \end{equation} 
\begin{figure}
\begin{center}
    \includegraphics[width=1.0\linewidth]{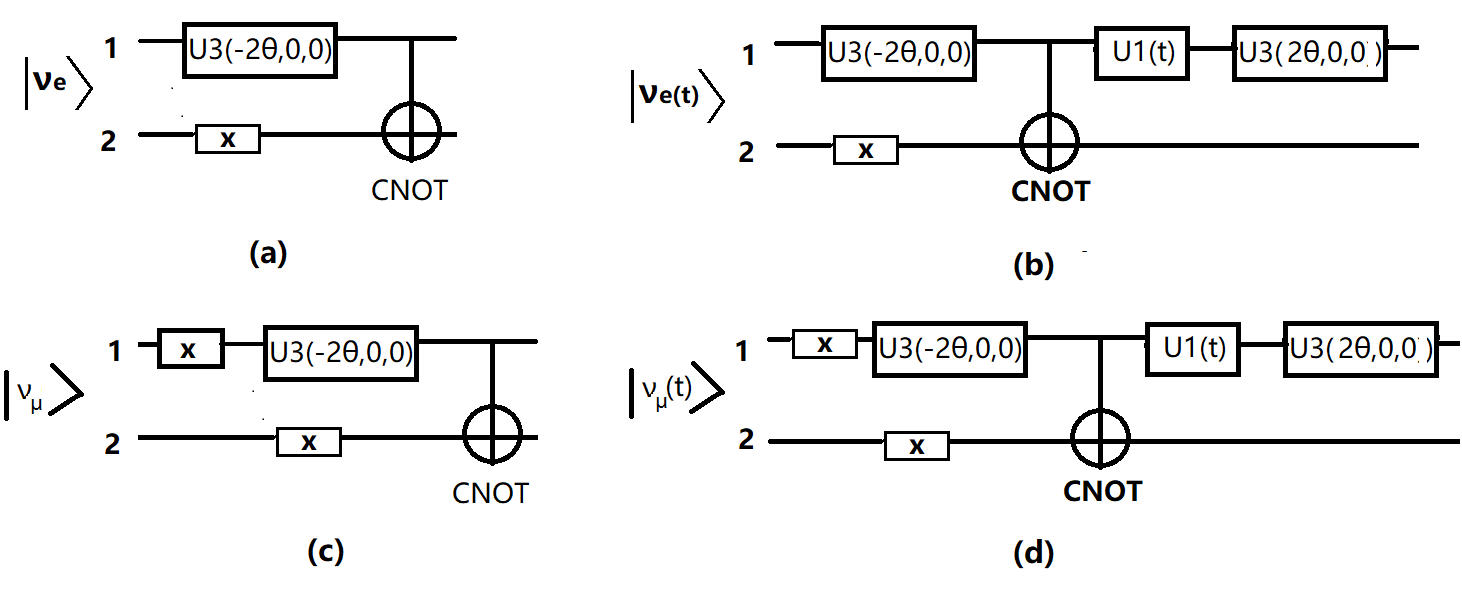}      
     \caption{  \textbf{(a)} The circuit represent an electron-flavor neutrino state in a linear superposition of mass mode basis in the two qubit system: $\ket{\nu_e}=\tilde{U}_{ee}\ket{10}_{1}+\tilde{U}_{e\mu}\ket{01}_{2}$. \textbf{(b)} The circuit represent the time evolved electron flavor neutrino state in a linear superposition of flavor mode basis ($\nu_e$ disappearance) in the two qubit system: $\ket{\nu_e(t)}=\tilde{U}_{ee}(t)\ket{10}_{e}+\tilde{U}_{e\mu}(t)\ket{01}_{\mu}$. \textbf{(c)} The circuit represent muon-flavor neutrino state in a linear superposition of mass mode basis in the two qubit system: $\ket{\nu_\mu}=\tilde{U}_{\mu e}\ket{10}_{1}+\tilde{U}_{\mu\mu}\ket{01}_{2}$. \textbf{(d)} The circuit represent the time evolved muon flavor neutrino state in a linear superposition of flavor mode basis ($\nu_\mu\rightarrow\nu_e$) in the two qubit system: $\ket{\nu_\mu(t)}=\tilde{U}_{\mu e}(t)\ket{10}_{e}+\tilde{U}_{\mu \mu}(t)\ket{01}_{\mu}$. Here, 1 and 2 represent the input qubits first and second, respectively.}
\end{center}    
 \end{figure}    
     
In order to calculate the concurrence we have to perform the spin-flipped operation on the density matrix. So to construct a quantum circuit to enable us to do this we have to prepare two  copies of bi-partite neutrino state $\ket{\nu_\alpha(t)}\otimes{\ket{\nu_\alpha (t)}}$ in the two flavor system (where $\alpha=e,\mu$), and apply a spin-flipped operation $\sigma_y\otimes\sigma_y$ on one of the two copies to prepare an arbitrary global state of neutrino in the four qubit system. The concurrence value of the time evolved flavor neutrino oscillation can be extracted from the global state \cite{Romero:2007}. In  Fig 4. the first two channels (1 and 2) stand for the entangled state $\ket{\nu_e(t)}$ that we want to measure. The third and fourth channel (3 and 4) denote the copy of $\ket{\nu_e(t)}$. Take two copies of the time evolved electron neutrino flavor state ($\nu_e(t)$): $\ket{\nu_e(t)}\otimes\ket{\nu_e(t)}$, and apply spin-flipped operation $\sigma_y\otimes\sigma_y$ on the second copy  by 
\begin{eqnarray}
&{\ket{\Phi(t)}}={\ket{\nu_e(t)}\otimes(\sigma_y\otimes\sigma_y\ket{\nu_e(t)}) =(\tilde{U}_{ee}(t)\ket{10}+\tilde{U}_{e\mu}(t)\ket{01})} {\otimes (\tilde{U}_{ee}(t)\ket{01}+\tilde{U}_{e\mu}(t)\ket{10})}&\nonumber \\
& ={(\tilde{U}_{ee}(t))^2\ket{1001}+\tilde{U}_{ee}(t)\tilde{U}_{e\mu}\ket{1010}+\tilde{U}_{e\mu}(t)\tilde{U}_{ee}\ket{0101} +(\tilde{U}_{e\mu}(t))^2\ket{0110}}&
\end{eqnarray}
Now  apply $CNOT_{24}$ operation between second (2) and fourth (4) qubit, and the target qubit (4) is inverted only when the control qubit (2) is $\ket{1}$ i.e, $\ket{0101}\rightarrow \ket{0100}$ and $\ket{0110}\rightarrow \ket{0111}$, such that 
\begin{eqnarray}
& {\ket{\Phi_1(t)}}={(\tilde{U}_{ee}(t))^2\ket{1001}+\tilde{U}_{ee}(t)\tilde{U}_{e\mu}\ket{1010}+\tilde{U}_{e\mu}(t)\tilde{U}_{ee}\ket{0100}
+(\tilde{U}_{e\mu}(t))^2\ket{0111}}&
\end{eqnarray}
Finally, we perform the Hadamard transformation, $H=\frac{1}{\sqrt{2}}\begin{pmatrix}
1 & 1\\
1 & -1\\
\end{pmatrix}$ on the second (2) qubit. The H operation can transfer each qubit as: $ H\ket{0}=\frac{1}{\sqrt{2}}(\ket{0}+\ket{1})$ and
    $H\ket{1}= \frac{1}{\sqrt{2}}(\ket{0}-\ket{1}).$ 
\begin{figure}
\begin{center}
     \includegraphics[width=1.0\linewidth]{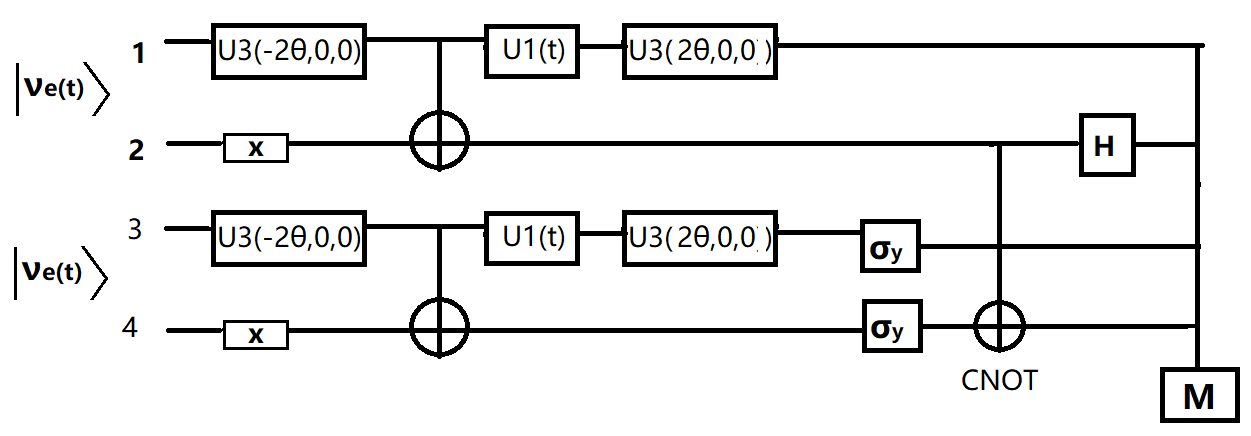}
     \caption{  The circuit represent the concurrence measurement of $\nu_e$ disappearance in two-flavor neutrino oscillations.}
\end{center}     
\end{figure}  
The state of the overall system is 
\begin{eqnarray}
&{\ket{\Phi_2(t)}}={\frac{1}{\sqrt{2}}[(\tilde{U}_{ee}(t))^2\ket{1001}-(\tilde{U}_{ee}(t))^2\ket{1101} + \tilde{U}_{ee}(t)\tilde{U}_{e\mu}(t)\ket{1010}-\tilde{U}_{ee}(t)\tilde{U}_{e\mu}(t)\ket{1110}}&\nonumber \\
&{+\tilde{U}_{e\mu}(t)\tilde{U}_{ee}(t)\ket{0100}+\tilde{U}_{e\mu}(t)\tilde{U}_{ee}(t)\ket{0000} +(\tilde{U}_{e\mu}(t))^2\ket{0111} 
+(\tilde{U}_{e\mu}(t))^2\ket{0011}].}&
\end{eqnarray}
The concurrence information of the electron neutrino flavor state $\ket{\nu_e(t)}$ is then the coefficient $\tilde{U}_{e\mu}(t)\tilde{U}_{ee}(t)$ and 
\begin{equation}
C(\ket{\nu_e(t)})=2\sqrt{2P_{0000}}=2\sqrt{P_{a}P_{d}},
\end{equation} where $P_{0000}=\frac{|\tilde{U}_{ee}(t)|^2|\tilde{U}_{e\mu}(t)|^2}{2}=\frac{P_{a}P_{d}}{2}$. The quantum simulation on IBMQ processor is shown in Fig5.

\begin{figure}
\begin{center}
     \includegraphics[width=1.0\linewidth]{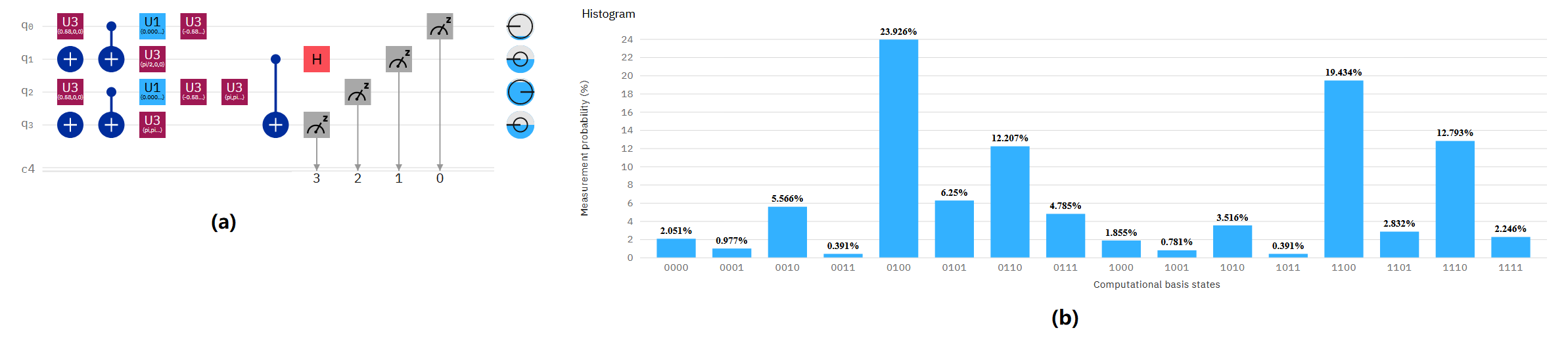} 
     \caption{{  \textbf{(a)} Concurrence circuit for the two qubit  $\nu_e$ disappearance bi-partite state on the IBMQ platform. \textbf{(b)} The concurrence varies with time at the IBMQ computer for an initial electron neutrino flavor state . The concurrence information  encoded in the coefficients of four qubit global state basis are shown through Histogram (probabilities in percentage) plot \cite{Esteban:2020cvm}.}}
\end{center}
\end{figure} 

 \section{Tri-Partite Entanglement In Three-Flavor Neutrino Oscillations}
The density matrix of the time evolved electron neutrino flavor state: $\ket{\nu_{e}(t)}=\tilde{U}_{ee}(t)\ket{100}_e+\tilde{U}_{e\mu}(t)\ket{010}_\mu+ \tilde{U}_{e\tau}(t)\ket{001}_{\tau}$, is \begin{eqnarray}
&{\rho^{e\mu\tau}(t) = \vert\tilde{U}_{ee}(t)\vert^2 \ket{100}\bra{100} + \vert\tilde{U}_{e\mu}(t)\vert^2\ket{010} \bra{010}+\vert\tilde{U}_{e\tau}(t)\vert^2 \ket{001}\bra{001} +\tilde{U}_{ee}(t)\tilde{U}_{e\mu}^{*}(t)}&\nonumber \\
&{ \ket{100}\bra{010}+\tilde{U}_{ee}(t) \tilde{U}_{e\tau}^{*}(t)\ket{100}\bra{001}+\tilde{U}_{e\mu}(t)\tilde{U}_{e\tau}^{*}(t)\ket{010}\bra{001} +\tilde{U}_{e\tau}(t)\tilde{U}_{ee}^{*}(t) }&\nonumber \\
&\hspace{-18em}{\ket{001}\bra{100}+\tilde{U}_{e\tau} (t)\tilde{U}_{e\mu}^{*}(t)\ket{001}\bra{010}.}&
\end{eqnarray}
 We calculate both pairwise entanglement and genuine tripartite entanglement. The pairwise measures of entanglement are similar to the bipartitite case because we have effectively reduced the three $\nu$ system to bi-partite system. These are  negativity ($N^2_{e(\mu\tau)}$), concurrence ($C^2_{e(\mu\tau)}$), tangle ($\tau_{e(\mu\tau)}$)
and linear entropy ($S_{e(\mu\tau))}$.  They are related as
    $N^2_{e(\mu\tau)}=C^2_{e(\mu\tau)}=\tau_{e(\mu\tau)}=S_{e(\mu\tau)}=4P_aP_d$.
For tri-partite entanglement an additional  criterion for entanglement is the Coffman-Kundu-Wooters (CKW) inequality. It states that the sum of quantum correlations between $e$ and $\mu$, and between $e$ and $\tau$, is either less than or equal to the quantum correlations between $e$ and $\mu\tau$ (treating it as a single object) \cite{OU:2007, Coffman:1999jd}: 
$C^{2} _{e\mu}+C^{2} _{e\tau}\leq C^{2} _{e(\mu\tau)}$ where  $C_{e\mu}$ and $C_{e\tau}$ are the concurrences of the mixed states $\rho^{e\mu}(t)=Tr_{\tau}(\rho^{e\mu\tau}(t))$ ,  $\rho^{e\tau}(t)=Tr_{\mu}(\rho^{e\mu\tau}(t))$ and $C_{e(\mu\tau)}=2\sqrt{det\rho^e(t)}$ with $\rho^e(t)=Tr_{\mu\tau}(\rho^{e\mu\tau}(t))$.
In the three neutrino case we find that 
$C^2 _{e\mu}+C^2 _{e\tau}=C^2 _{e(\mu\tau)}$, and is unchanged by permutations .
 The other measures of tripartite entanglement are the tangle equality $\tau_{e\mu}+\tau_{e\tau}=\tau_{e(\mu\tau)}$ and the  negativity inequality  $N^{2} _{e\mu}+N^{2} _{e\tau}< N^{2} _{e(\mu\tau)}$.
There are two extra measure for genuine tri-partite entanglement quantified by three-tangle and three-$\pi$ negativity known as residual entanglement. 
The three-tangle is   $\tau_{e\mu\tau}=C^{2} _{e(\mu\tau)}-C^{2} _{e\mu}-C^{2} _{e\tau}=0$, but there is  non-zero value of the residual entanglment  three-$\pi$ given by  $\pi_{e\mu\tau}=\frac{1}{3}(N^2 _{e(\mu\tau)}+N^2 _{\mu(e\tau)} +N^2 _{\tau(e\mu)}-2N^2 _{e\mu}-2N^2 _{e\tau}-2N^2 _{\mu\tau})>0$ (see Fig.6(a)) .
This shows that the three flavor neutrino oscillations has a genuine form of tri-partite entanglement. The result also shows that the correlation exhibited by neutrino oscillations in tri-partite system are like the W-states which are legitimate physical resources for quantum information tasks \cite{Jha:2020hyh}. 

\begin{figure}
\begin{center}
     \includegraphics[width=1.0\linewidth]{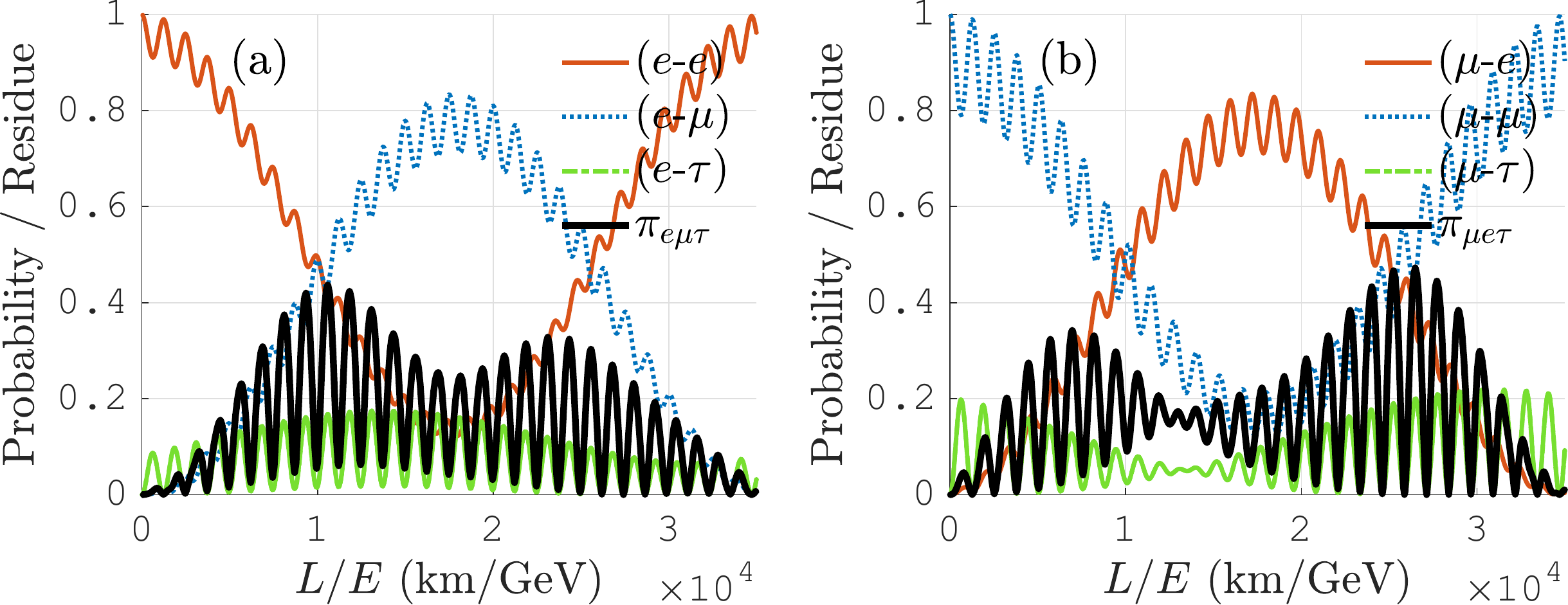}
     \caption{\textbf{(a).} Time evolved electron neutrino flavor state $\ket{\nu_e(t)}$ (relevant to reactor experiment) and \textbf{(b)} a muon flavor state $\ket{\nu_{\mu}(t)}$ (relevant to accelerator experiment) vs scale of distance per energy unit $\frac{L}{E}$. At  $\frac{L}{E}>0$ entanglement among three-flavor modes occurs i.e, the black curve $\pi_{e\mu\tau}>0$ or $\pi_{\mu e\tau}>0$, and exhibits a typical oscillatory behavior. $\pi_{e\mu\tau}$ reaches the maximum value 0.436629 (see Fig.(a)) when transition probabilities are $P_{\nu_{e\rightarrow e}}=0.39602$, $P_{\nu_{e\rightarrow \mu}}=0.435899$, and  $P_{\nu_{e\rightarrow\tau}}=0.168081$. Similarly, for $\ket{\nu_{\mu}(t)}$, $\pi_{\mu e \tau}$ reaches the maximum value 0.472629 (see Fig.(b)) indicating genuine tri-partite entanglement \cite{Esteban:2018azc}}.
\end{center}     
\end{figure}
 \section{Conclusion} 
 We have quantified  bipartite measures like concurrence, tangle, negativity and linear entropy in two and three-flavor neutrino oscillations.  We  have constructed quantum computer circuit using Universal $U(3)$ gate, S-gate, Controlled-NOT and Pauli (X) gate to outline the simulation of two flavor neutrino oscillations on a quantum computer. We directly measure the concurrence using spin-flipped $\sigma_y\otimes\sigma_y$ gate, and Hadamard gate, and proposed the implications of the implementation of entanglement in the two neutrino system on the IBM quantum processor. The tri-partite result for residual entanglement $\pi_{\mu e\tau}>0$ imply that the three-neutrino state has a  genuine form of three way entanglement. Quantum simulation of tri-partite entanglement for the three neutrino system on a quantum computer is in progress and will be reported a more  comprehensive paper.  The authors acknowledge the use of IBM Quantum computer for this work. The views expressed are those of the authors and do not reflect the official policy or position of IBM or the IBMQ team.



\section{References}
 \begin{thebibliography}{9}
\bibitem{Jha:2020hyh}
A.~K.~Jha, S.~Mukherjee and B.~A.~Bambah,
[arXiv:2004.14853 [hep-ph]].

\bibitem{Blasone:2007vw} 
  M.~Blasone, F.~Dell'Anno, S.~De Siena and F.~Illuminati,
  EPL {\bf 85}, 50002 (2009)
  doi:10.1209/0295-5075/85/50002
  [arXiv:0707.4476 [hep-ph]].
 
\bibitem{Arguelles:2019phs}
C.~A.~Arg\"uelles and B.~J.~P.~Jones,
Phys. Rev. Research. \textbf{1} (2019), 033176
doi:10.1103/PhysRevResearch.1.033176
[arXiv:1904.10559 [quant-ph]].
 
 
  
\bibitem{DiMolfetta:2016gzc}
G.~Di Molfetta and A.~P\'erez,
New J. Phys. \textbf{18} (2016) no.10, 103038
doi:10.1088/1367-2630/18/10/103038
[arXiv:1607.00529 [quant-ph]].
  
\bibitem{DARIUSZ:2012}
DARIUSZ KURZYK,``Introduction to Quantum entanglement,'' Theoretic and Applied Informatics {\bf 24}, IISN 1896-5334, Vol.24(2020),no.2,pp.135-150. DOI:10.2478/v10179-012-0010-7




\bibitem{OU:2007}
Yong-Cheng Ou, Heng Fan, 
Physical Review A 75, 062308 (2007), doi:10.1103/PhysRevA.75.062308, [ arXiv:quant-ph/0702127v4]

  
\bibitem{Coffman:1999jd} 
  V.~Coffman, J.~Kundu and W.~K.~Wootters,
  Phys.\ Rev.\ A {\bf 61}, 052306 (2000)
  doi:10.1103/PhysRevA.61.052306
  [quant-ph/9907047].
  %
  
\bibitem{Esteban:2018azc}
I.~Esteban, M.~C.~Gonzalez-Garcia, A.~Hernandez-Cabezudo, M.~Maltoni and T.~Schwetz,
JHEP \textbf{01} (2019), 106
doi:10.1007/JHEP01(2019)106
[arXiv:1811.05487 [hep-ph]].
  
\bibitem{Esteban:2020cvm}
I.~Esteban, M.~C.~Gonzalez-Garcia, M.~Maltoni, T.~Schwetz and A.~Zhou,
JHEP \textbf{09} (2020), 178
doi:10.1007/JHEP09(2020)178
[arXiv:2007.14792 [hep-ph]].

\bibitem{An:2015rpe}
F.~P.~An \textit{et al.} [Daya Bay],
Phys. Rev. Lett. \textbf{115} (2015) no.11, 111802
doi:10.1103/PhysRevLett.115.111802
[arXiv:1505.03456 [hep-ex]].



\bibitem{Sousa:2015bxa}
A.~B.~Sousa [MINOS and MINOS+],
AIP Conf. Proc. \textbf{1666} (2015) no.1, 110004
doi:10.1063/1.4915576
[arXiv:1502.07715 [hep-ex]].



\bibitem{Romero:2007}
 G. Romero, C.E. Lopez, F. Lastra, E. Solano, J.C. Retamal, Direct measurement of concurrence for atomic two-qubit pure states, Phys. Rev. A 75, 032303 (2007), arXiv:quant-ph/0611016v1
 
 
  
\end{thebibliography}
\end{document}